\documentstyle[aps,epsfig]{revtex}
\newcommand{\lsim}{\mathrel{\rlap{\lower4pt\hbox{\hskip0pt$\sim$}}
\raise1pt\hbox{$<$}}}
\newcommand{\gsim}{\mathrel{\rlap{\lower4pt\hbox{\hskip0pt$\sim$}}
\raise1pt\hbox{$>$}}}
\newcommand{\sfrac}[2]{\mbox{\footnotesize $\frac{#1}{#2}$}}
\begin{document}
\twocolumn

\title{Nucleon form factors and a nonpointlike diquark}

%
\author{J.C.R. Bloch,\footnotemark[1] C.D. Roberts,\footnotemark[1]
S.M. Schmidt,\footnotemark[1] A. Bender\footnotemark[2] and
M.R. Frank\footnotemark[3]\vspace*{0.2em}}
\address{\footnotemark[1]Physics Division, Building 203, Argonne
National Laboratory, Argonne IL 60439-4843\\[0.1em]
\footnotemark[2]Centre for the Subatomic Structure of Matter, University of
Adelaide, Adelaide SA 5005, Australia\\[0.1em]
\footnotemark[3]Institute for Nuclear Theory, University of Washington, 
Seattle, WA 98195}
\date{Preprint: ANL-PHY-9382-TH-99; Pacs Numbers: 24.85.+p, 13.40.Gp,
14.20.Dh, 12.38.Lg}
\maketitle

\vspace*{-1.6em}

\begin{abstract}
Nucleon form factors are calculated on $q^2\in[0,3]\,$GeV$^2$ using an {\it
Ansatz} for the nucleon's Fadde'ev amplitude motivated by quark-diquark
solutions of the relativistic Fadde'ev equation.  Only the scalar diquark is
retained, and it and the quark are confined.  A good description of the data
requires a nonpointlike diquark correlation with an electromagnetic radius of
$0.8\,r_\pi$.  The composite, nonpointlike nature of the diquark is crucial.
It provides for diquark-breakup terms that are of greater importance than the
diquark photon absorption contribution.
\end{abstract}\hspace*{0.4em}

%
%
%
%
%
%
%
%
Mesons present a two-body problem, and the Dyson-Schwinger equations (DSEs)
have been widely used in the calculation of their properties and
interactions~\cite{cdragw,echaya}.  Many studies have focused on
electromagnetic processes; such as the form factors of light
pseudoscalar~\cite{mark,mr98} and vector mesons~\cite{mikeb}, and the
$\gamma^\ast \pi^0\to\gamma$~\cite{cdrcroat,pctcroat,dubravko},
$\gamma^\ast\pi\to\rho$~\cite{pctcroat} and
$\gamma\pi^\ast\to\pi\pi$~\cite{ar96} transition form factors, all of which
are accessible at TJNAF.  These studies provide a foundation for the
exploration of nucleons, which is fundamentally a three-body problem.

The nucleon's bound state amplitude can be obtained from a relativistic
Fadde'ev equation~\cite{rega}.  Its analysis may be simplified by using the
feature that ladder-like dressed-gluon exchange between quarks is attractive
in the colour antitriplet channel.  Then, in what is an analogue of the
rainbow-ladder truncation for mesons, the Fadde'ev equation can be reduced to
a sum of three coupled equations, in which the primary dynamical content is
dressed-gluon exchange generating a correlation between two quarks and the
iterated exchange of roles between the dormant and diquark-participant
quarks.  Following this approach, the diquark correlation is represented by
the solution of an homogeneous Bethe-Salpeter equation in the dressed-ladder
truncation and hence its contribution to the quark-quark scattering matrix,
${\cal M}_{qq}$, is that of an asymptotic bound state; i.e., it contributes a
simple pole.  That is an artefact of the ladder truncation~\cite{brs96} and
complicates solving the Fadde'ev equation~\cite{bcp89} by introducing
spurious free-particle singularities in the kernel.

Studies of DSE-models~\cite{cdragw,echaya} suggest that confinement
can be realised via the absence of a Lehmann representation for coloured
Green functions, and have led to a phenomenologically efficacious
parametrisation of the dressed-quark Schwinger function~\cite{mark}.  A
similar parametrisation of the diquark contribution to ${\cal M}_{qq}$,
advocated in Ref.~\cite{bsesep}, has been used to good effect in solving the
Fadde'ev equation~\cite{raA}.  We use such representations herein.

The nucleon-photon current is\footnote{In our Euclidean formulation: $p\cdot
q=\sum_{i=1}^4 p_i q_i$, $\{\gamma_\mu,\gamma_\nu\}=2\,\delta_{\mu\nu}$,
$\gamma_\mu^\dagger = \gamma_\mu$, $\sigma_{\mu\nu}=
\sfrac{i}{2}[\gamma_\mu,\gamma_\nu]$, and
tr$_D[\gamma_5\gamma_\mu\gamma_\nu\gamma_\rho\gamma_\sigma]=
-4\,\epsilon_{\mu\nu\rho\sigma}$, $\epsilon_{1234}= 1$.}
\begin{eqnarray}
\label{Jnucleon}
J_\mu(P^\prime,P) & = & ie\,\bar u(P^\prime)\, \Lambda_\mu(q,P) \,u(P)\,,
\end{eqnarray}
where the spinors satisfy: $\gamma\cdot P \, u(P) = i M u(P)$, $\bar u(P)\,
\gamma\cdot P = i M \bar u(P)$, with $M=0.94\,$GeV the nucleon mass, and
$q=(P^\prime-P)$.  The complete specification of a fermion-vector-boson
vertex requires twelve independent scalar functions:
\begin{eqnarray}
\nonumber 
\lefteqn{i\Lambda_\mu(q,P)  =   i \gamma_\mu\,f_1
+ i\sigma_{\mu\nu}\,q_\nu\,f_2
+ R_\mu\,f_3 
+ i\gamma\cdot R\,R_\mu\,f_4
}\\
&& 
+ i\sigma_{\nu\rho}\,R_\mu\,q_\nu\,R_\rho\,f_5
+ i\gamma_5\gamma_\nu\,\varepsilon_{\mu\nu\rho\sigma}\,
         q_\rho\,R_\sigma\,f_6\,+ \ldots\,,
\end{eqnarray}
where $f_i= f_i(q^2,q\cdot P,P^2)$, $R= (P^\prime + P)$ and $q\cdot R = 0$
for elastic scattering.  However, using the definition of the nucleon
spinors, (\ref{Jnucleon}) can be written
\begin{eqnarray}
\label{F1F2}
\lefteqn{J_\mu(P^\prime,P) =}\\
&&\nonumber
ie\,\bar u(P^\prime)\left(\gamma_\mu \, F_1(q^2) 
+ \frac{1}{2 M}\, \sigma_{\mu\nu}\, q_\nu\, F_2(q^2)\right)u(P)\,,
\end{eqnarray}
where the Dirac and Pauli form factors are
\begin{eqnarray}
\lefteqn{F_1  = }\\
&& \nonumber f_1 
+ 2\,M\,f_3 - 4\, M^2\,f_4 - 2\,M\,q^2\,f_{5} - q^2\,f_{6}\,,\\
\lefteqn{F_2  = }\\
&& \nonumber  2\,M\,f_2
-2\,M\,f_3 + 4\,M^2\,f_4 + 2\,M\,f_{5} - 4\,M^2\,f_{6}\,,
\end{eqnarray}
in terms of which one has the electric and magnetic form factors:
\begin{eqnarray}
G_E(q^2) & = & F_1(q^2) - \frac{q^2}{4 M^2} F_2(q^2)\,,\\
G_M(q^2) & = & F_1(q^2) + F_2(q^2)\,.
\end{eqnarray}

%
To calculate these form factors we represent the nucleon as a three-quark
bound state involving a diquark correlation, and require the photon to probe
the diquark's internal structure.  Antisymmetrisation ensures there is an
exchange of roles between the dormant and diquark-participant quarks and this
gives rise to diquark ``breakup'' contributions.  We describe the propagation
of the dressed-quarks and diquark correlation by confining parametrisations
and hence pinch singularities associated with quark production thresholds are
absent.  Our calculation is kindred to many studies of meson
properties~\cite{mark,mr98,mikeb,cdrcroat,pctcroat,ar96}.

We write the Fadde'ev amplitude of the nucleon as~\cite{axel}
\begin{eqnarray}
\label{Psi}
\lefteqn{\Psi^{\tau}_\alpha(p_1,\alpha_1,\tau^1;p_2,\alpha_2,\tau^2;
                p_3,\alpha_3,\tau^3)   = } \\
&& \nonumber 
\varepsilon_{c_1 c_2 c_3}\,
\delta^{\tau \tau^3}\,\delta_{\alpha \alpha_3}\,\psi(p_1+p_2,p_3)\,
\Delta(p_1+p_2)\,
\Gamma_{\alpha_1 \alpha_2}^{\tau^1 \tau^2}(p_1,p_2) \,,
\end{eqnarray}
where $\varepsilon_{c_1 c_2 c_3}$ effects a singlet coupling of the quarks'
colour indices, $(p_i,\alpha_i,\tau^i)$ denote the momentum and the Dirac and
isospin indices for the $i$-th quark constituent, $\alpha$ and $\tau$ are
these indices for the nucleon itself,
$\psi(\ell_1,\ell_2)$
is a Bethe-Salpeter-like amplitude characterising the relative-momentum
dependence of the correlation between diquark and quark, $\Delta(K)$
describes the propagation characteristics of the diquark, and
\begin{eqnarray}
\label{gdq}
\Gamma_{\alpha_1 \alpha_2}^{\tau^1 \tau^2}(p_1,p_2) & = &
(C i\gamma_5)_{\alpha_1 \alpha_2}\, (i\tau_2)^{\tau^1\tau^2}\,
\Gamma (p_1,p_2)
\end{eqnarray}
represents the momentum-dependence, and spin and isospin character of the
diquark correlation; i.e., it corresponds to a diquark Bethe-Salpeter
amplitude.

With this form of $\Psi$, we retain in ${\cal M}_{qq}$ only the contribution
of the scalar diquark, which has the largest correlation
length~\cite{bsesep}: $\lambda_{0^+}:=1/m_{0^+} = 0.27\,$fm.  For all
$(ud)$-correlations with $J^P\neq 1^+$, $\lambda_{ud} < 0.5\,\lambda_{0^+}$.
The axial-vector correlation is different: $\lambda_{1^+} =
0.78\,\lambda_{0^+}$, and it is quantitatively important in the calculation
of baryon masses ($\lsim 30$\%)~\cite{raA}.  Hence we anticipate that
neglecting the $1^+$ correlation will prove the primary defect of our {\it
Ansatz}.  However, it is an helpful expedient in this exploratory
calculation, which is made complicated by our desire to elucidate the effect
of the diquarks' internal structure.

\begin{figure}
\epsfig{figure=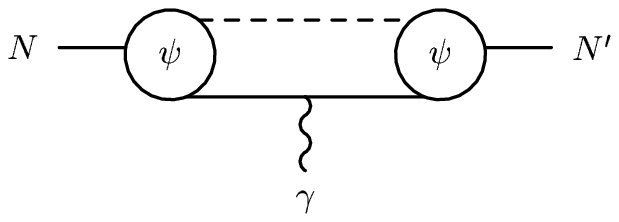,height=1.5cm}\vspace*{-3.8\baselineskip}

\hspace*{\fill}\epsfig{figure=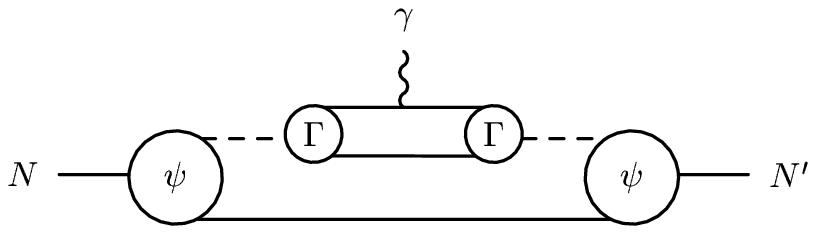,height=1.5cm}

\epsfig{figure=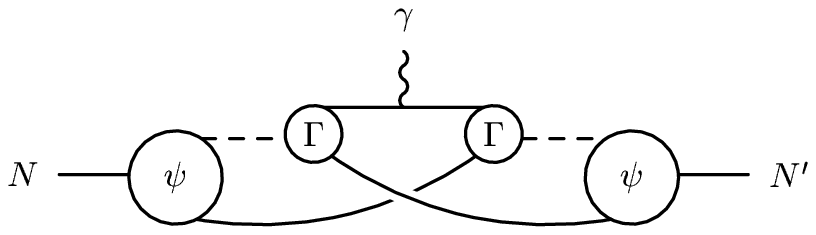,height=1.5cm}\vspace*{0.6\baselineskip}

\hspace*{\fill}\epsfig{figure=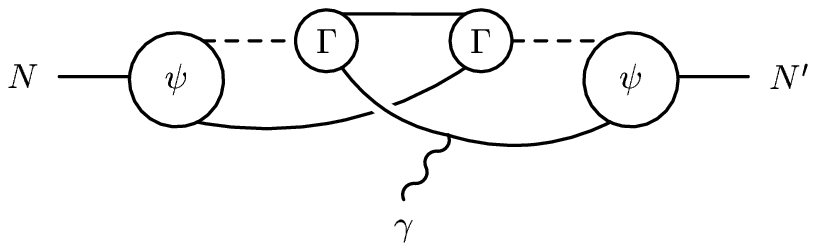,height=1.5cm}\vspace*{-1.2\baselineskip}

\epsfig{figure=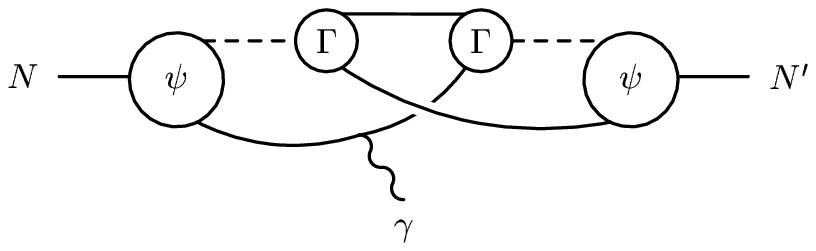,height=1.5cm}
\caption{Our impulse approximation to the electromagnetic current requires
the calculation of five contributions, (\protect\ref{L1}) --
(\protect\ref{L5}).  $\psi$: $\psi(\ell_1,\ell_2)$ in (\protect\ref{Psi});
$\Gamma$: Bethe-Salpeter-like diquark amplitude in (\protect\ref{gdq}); solid
line: $S(q)$, quark propagator in (\protect\ref{qprop}); dotted line:
$\Delta(K)$, diquark propagator in (\protect\ref{dprop}).  The lowest three
diagrams, which describe the interchange between the dormant quark and the
diquark participants, effect the antisymmetrisation of the nucleon's Fadde'ev
amplitude.  Current conservation follows because the photon-quark vertex is
dressed, given in (\protect\ref{bcvtx}).
\label{ff}}
\end{figure}

Our impulse approximation to the nucleon form factor is depicted in
Fig.~\ref{ff}.  Enumerating from top to bottom, the diagrams represent
\begin{eqnarray}
\label{L1}
\lefteqn{\Lambda^1_\mu(q,P)  =
3 \int\!\!\sfrac{d^4 \ell}{(2\pi)^4}\,
}\\
& & \nonumber 
\psi(K,p_3+q) \Delta(K) \psi(K,p_3)\,Q_F\Lambda_\mu^q(p_3+q,p_3) \,,
\end{eqnarray}
with\footnote{$\eta$ describes the partitioning of the nucleon's total
momentum: $P= p_1+p_2+p_3$, between the diquark and quark, a necessary
feature of a covariant treatment.}
$K= \eta P + \ell$, $p_3= (1-\eta) P -\ell$, $p_2= K/2 - k$, $Q_F={\rm
diag}(2/3,-1/3)$,
$\Lambda_\mu^q(k_1,k_2) = S(k_1)\,\Gamma_\mu(k_1,k_2)\,S(k_2)$, 
\begin{eqnarray}
\lefteqn{\Lambda^2_\mu(q,P) =
6\int\!\!\sfrac{d^4 k}{(2\pi)^4}\sfrac{d^4 \ell}{(2\pi)^4}\,
\Omega(p_1+q,p_2,p_3)\, }\\
&& \nonumber
\times \Omega(p_1,p_2,p_3) \,
{\rm tr}_D\left[\Lambda_\mu^q(p_1+q,p_1) S(p_2)\right]\,
S(p_3)\,\sfrac{1}{3} I_F\,,
\end{eqnarray}
which contributes equally to the proton and neutron and contains the diquark
electromagnetic form factor, with $6 = \varepsilon_{c_1 c_2 c_3}
\varepsilon_{c_1 c_2 c_3}$ and
\begin{equation}
\Omega(p_1,p_2,p_3) = \psi(p_1+p_2,p_3) \,\Delta(p_1+p_2) \, 
\Gamma(p_1,p_2)\,,
\end{equation}
\begin{eqnarray}
\lefteqn{\Lambda^3_\mu(q,P) = 6\int\!\!\sfrac{d^4 k}{(2\pi)^4}\sfrac{d^4
\ell}{(2\pi)^4}\,\Omega(p_1+q,p_3,p_2)\,   }\\
&& \nonumber 
\times \Omega(p_1,p_2,p_3)\,
S(p_2) \, (i\tau_2)^{\rm T} Q_F(i \tau_2)\,
\Lambda^q_\mu(p_1,p_1+q)\,S(p_3)\,,\\
\lefteqn{\Lambda^4_\mu(q,P) = 6\int\sfrac{d^4 k}{(2\pi)^4}\sfrac{d^4
\ell}{(2\pi)^4}\,\Omega(p_1,p_3,p_2+q)\,  }\\
&& \nonumber 
\times \Omega(p_1,p_2,p_3) \,
Q_F\Lambda^q_\mu(p_2+q,p_2) \,S(p_1) \, S(p_3)\,,\\
\label{L5}
\lefteqn{\Lambda^5_\mu(q,P) = 6\int\sfrac{d^4 k}{(2\pi)^4}\sfrac{d^4
\ell}{(2\pi)^4}\, \Omega(p_1,p_3+q,p_2)\,  }\\
&& \nonumber 
\times \Omega(p_1,p_2,p_3)\,
S(p_2)\,S(p_1)\,Q_F\Lambda_\mu(p_3+q,p_3) \,.
\end{eqnarray}
The nucleon-photon vertex is
\begin{eqnarray}
\label{nucvtx}
\Lambda_\mu(q,P) &=&
\Lambda^1_\mu(q,P)
+ 2 \sum_{i=2}^5\,\Lambda^i_\mu(q,P)\,.
\end{eqnarray}

(\ref{nucvtx}) is fully defined once $\Psi \sim \psi\,\Gamma\,\Delta$, $S$
and $\Gamma_\mu$ are specified.  $S$ and $\Gamma_\mu$ are primary elements in
studies of meson properties and are already well constrained.  For the
dressed-quark propagator:
\begin{eqnarray}
\label{qprop}
S(p) & = & -i\gamma\cdot p\, \sigma_V(p^2) + \sigma_S(p^2)\\
& = & \left[i \gamma\cdot p \, A(p^2) + B(p^2)\right]^{-1}\,,
\end{eqnarray}
we use the algebraic parametrisations~\cite{mark}: 
\begin{eqnarray}
\label{ssm}
\bar\sigma_S(x) & =&  2\,\bar m \,{\cal F}(2 (x+\bar m^2))\\
&& \nonumber
+ {\cal F}(b_1 x) \,{\cal F}(b_3 x) \,
\left[b_0 + b_2 {\cal F}(\epsilon x)\right]\,,\\
\bar\sigma_V(x) & = & \frac{1}{x+\bar m^2}\,
\left[ 1 - {\cal F}(2 (x+\bar m^2))\right]\,,
\end{eqnarray}
with ${\cal F}(y) = (1-{\rm e}^{-y})/y$, $x=p^2/\lambda^2$, $\bar m$ =
$m/\lambda$, 
$\bar\sigma_S(x)  =  \lambda\,\sigma_S(p^2)$   
and $\bar\sigma_V(x)  =  \lambda^2\,\sigma_V(p^2)$. 
The mass-scale, $\lambda=0.566\,$GeV, and parameter values
\begin{equation}
\label{tableA} 
\begin{array}{ccccc}
   \bar m& b_0 & b_1 & b_2 & b_3 \\\hline
   0.00897 & 0.131 & 2.90 & 0.603 & 0.185 \;\;,
\end{array}
\end{equation}
were fixed in a least-squares fit to light-meson observables.
($\epsilon=10^{-4}$ in (\ref{ssm}) acts only to decouple the large- and
intermediate-$p^2$ domains.)  This algebraic parametrisation combines the
effects of confinement and DCSB with free-particle behaviour at large
spacelike $p^2$~\cite{echaya}.

In (\ref{L1})--(\ref{L5}), $\Gamma_\mu$ is the dressed-quark-photon vertex.
It satisfies the vector Ward-Takahashi identity:
\begin{equation}
\label{vwti}
(\ell_1 - \ell_2)_\mu \, i\Gamma_\mu(\ell_1,\ell_2) = 
S^{-1}(\ell_1) - S^{-1}(\ell_2)\,,
\end{equation}
which ensures current conservation~\cite{mark}.  $\Gamma_\mu$ has been much
studied~\cite{ayse97} and, although its exact form remains unknown, its
qualitative features have been elucidated so that a phenomenologically
efficacious {\it Ansatz} has emerged~\cite{bc80}:
\begin{eqnarray}
\label{bcvtx}\lefteqn{i\Gamma_\mu(\ell_1,\ell_2) = 
i\Sigma_A(\ell_1^2,\ell_2^2)\,\gamma_\mu }\\ 
& & \nonumber 
+
(\ell_1+\ell_2)_\mu\,\left[\sfrac{1}{2}i\gamma\cdot (\ell_1+\ell_2) \,
\Delta_A(\ell_1^2,\ell_2^2) + \Delta_B(\ell_1^2,\ell_2^2)\right]\,;\\
&&  \Sigma_F(\ell_1^2,\ell_2^2) = \sfrac{1}{2}\,[F(\ell_1^2)+F(\ell_2^2)]\,,\\
&& \Delta_F(\ell_1^2,\ell_2^2) =
\frac{F(\ell_1^2)-F(\ell_2^2)}{\ell_1^2-\ell_2^2}\,,
\end{eqnarray}
where $F= A, B$; i.e., the scalar functions in (\ref{qprop}).  A feature of
(\ref{bcvtx}) is that $\Gamma_\mu$ is completely determined by the
dressed-quark propagator.  Further, we estimate that calculable improvements
would modify our results by $\lsim 15\,$\%~\cite{piloop}.

The new element herein is the model of the nucleon's Fadde'ev amplitude,
(\ref{Psi}).  For the Bethe-Salpeter-like amplitudes we use the one-parameter
model forms
\begin{eqnarray}
\label{gammaell}
\Gamma(q_1,q_2) & = & 
 \frac{1}{{\cal N}_\Gamma}\,
        {\cal F}(q^2/\omega_\Gamma^2)\,,\;
q:= \sfrac{1}{2}(q_1-q_2)\\
\label{psiell}
\psi(\ell_1,\ell_2) & = & \frac{1}{{\cal N}_\Psi}\,{\cal
F}(\ell^2/\omega_\psi^2)\,,\;\ell := (1-\eta)\,\ell_1 - \eta\,\ell_2\,.
\end{eqnarray}
Our impulse approximation is founded on a dressed-ladder kernel in the
Fad\-de'ev equation and $\Gamma_\mu$ satisfies (\ref{vwti}).  Hence, the
canonical normalisation conditions for the diquark and nucleon amplitudes
translate to the constraints that the $(ud)$-diquark must have charge $1/3$
and the proton unit charge, which fix ${\cal N}_\Gamma$ and ${\cal N}_\Psi$.
For the diquark propagator we use the one-parameter form
\begin{equation}
\label{dprop}
\Delta(K^2)  =  \frac{1}{m_\Delta^2}\,{\cal F}(K^2/\omega_\Gamma^2)\,,
\end{equation}
and interpret $1/m_\Delta$ as the diquark correlation length.

%
We fix the model's three parameters by optimising a fit to $G^p_E(q^2)$ and
ensuring $G_E^n(0)=0$, which yields\footnote{
Our results are sensitive to $\eta$ because (\protect\ref{gammaell}) and
(\protect\ref{psiell}) are equivalent to retaining only the leading Dirac
amplitude in the expression for these functions and neglecting their $q\cdot
K$, $\ell\cdot P$ dependence when solving the Bethe-Salpeter and Fadde'ev
equations.  $\eta=2/3$ is required for this {\it Ansatz} to transform
correctly under charge conjugation.  Accounting for the $q\cdot K$,
$\ell\cdot P$ dependence would eliminate this
artefact~\protect\cite{raA,mr97}.  }
\begin{equation}
\label{params}
\begin{array}{l|ccc}
          & \;\omega_\psi & \omega_\Gamma & m_\Delta \\\hline
\eta=2/3\;& \;0.20  & 1.0   & 0.63
\end{array}
\end{equation}
all in GeV ($1/m_\Delta = 0.31\,$fm). Using Monte-Carlo methods to evaluate
the multi-dimensional integrals, these values give
\begin{equation}
\label{results}
\begin{array}{l|c|cc}
            & {\rm emp.}
                        & {\rm calc.} \\\hline       
r_p^2 \,({\rm fm})^2 &  (0.87 )^2       
                        & (0.79)^2 \\
r_n^2\, ({\rm fm})^2 & -(0.34 )^2\;\;\; 
                        & -(0.43)^2\;\;\; \\
\mu_p\, (\mu_N)      &  2.79 
                        & 2.88 \\
\mu_n\, (\mu_N)      & -1.91 \;\;\;
                        & -1.58\;\;\; \\     
\mu_n/\mu_p        & -0.68\;\;\;
                        & -0.55\;\;\; 
\end{array}
\end{equation}
where the statistical error is $\lsim 1\,$\%.  The sensitivity of our results
to the model's parameters is illustrated in Table~\ref{sens}.  It is clear
that the fit is stable but does not bracket the experimental domain; i.e.,
the model lacks a relevant degree of freedom, a defect we expect including an
axial-vector diquark to ameliorate.

The charge radii are obtained via
\begin{eqnarray}
r_{p,n}^2 
& = & -6 \left.\frac{d}{dq^2}\,F_1^{p,n}(q^2)\right|_{q^2=0} 
+ \frac{3}{2 M^2} F_2^{p,n}(0)\,,\\
& := & (r^I_{p,n})^2 + (r^F_{p,n})^2 
\end{eqnarray}
and in this calculation (in fm$^2$) 
\begin{equation}
\begin{array}{ll}
(r_p^I)^2 = \;\;\,(0.70)^2\,,\; & (r_p^F)^2 = \;\;\,(0.35)^2 \,,\\
(r_n^I)^2 = -(0.29)^2\,,\; & (r_n^F)^2 = -(0.32)^2 \,.
\end{array}
\end{equation}
A $20$\% reduction in $\omega_\Gamma$ (Table~\ref{sens}, row 4) reduces
$|r_n|$ by $7$\%.  However, that results from a $21$\% reduction in $|r_n^I|$
and $2$\% increase in $|r_n^F|$.  We attribute our overestimate of $|r_n^2|$
to a poor description of $F_1^n(q^2)$, which involves many cancellations
between terms because of the $(u,d,d)$ electric charge combinations and must
vanish at $q^2=0$.

Five diagrams contribute to our impulse approximation and diagram~2 involves
the diquark form factor.  The calculated value of the associated elastic
charge radius provides a measure of the size of the ``constituent'' diquark:
\begin{eqnarray}
r_{\rm 0^+}^2 & = & (0.45\,{\rm fm})^2 = (0.80\,r_\pi)^2\,,
\end{eqnarray}
with $r_\pi$ calculated in the same model~\cite{mark}, and in quantitative
agreement with another estimate~\cite{raB}.  This is important because, with
$\omega_\Gamma$ allowed to vary, $r_{0^+}$ is a qualitative prediction of the
model.  Thus an optimal description of the data {\it requires} a nonpointlike
diquark.

Table~\ref{ff1to5} provides a guide to each diagram's relative importance.
In all cases the first diagram, describing scattering from the dormant quark,
is the most significant.  For the charge radii the breakup contributions are
comparable in magnitude to the second diagram, photon-diquark scattering.
The magnetic moments are of particular interest.  A scalar diquark does not
have a magnetic moment, and that is expressed in our calculation by the very
small contribution from diagram $2$.  It is not identically zero because of
the {\it confinement} of the spectator quark; i.e., the absence of a
mass-shell.  Diagrams $3$-$5$ only appear because the diquark is a
nonpointlike composite and they provide $\sim 50\,$\% of $\mu_p$, $\mu_n$.
Discarding these contributions one obtains $\mu_n/\mu_p\geq -0.5$, and in
pointlike diquark models the axial-vector has alone been forced to remedy
that defect~\cite{keinerB}.  Our results indicate that approach to be
erroneous, attributing too much importance to the axial-vector correlation.

\begin{figure}[t]
\centering{\ \epsfig{figure=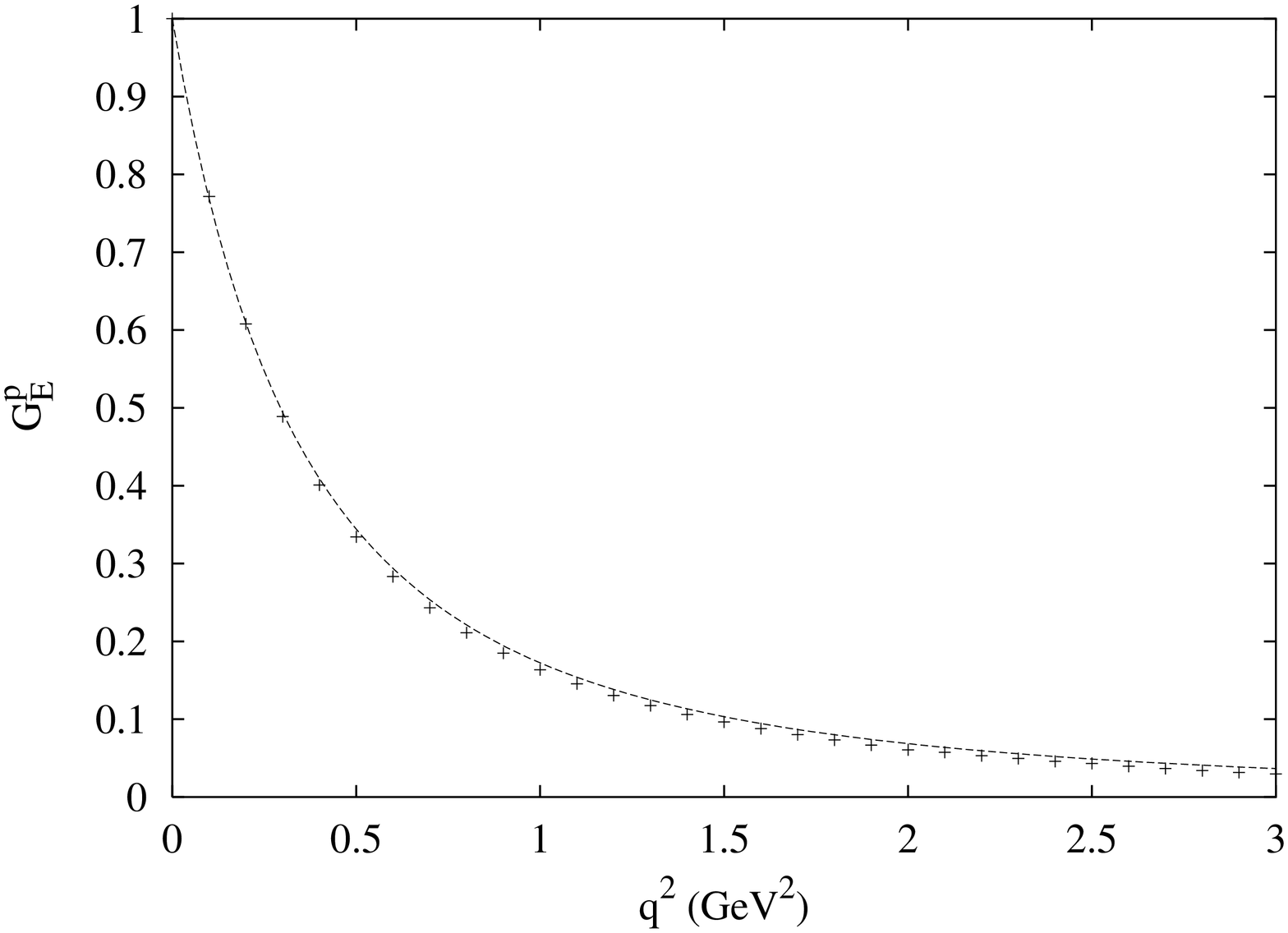,height=8.0cm,angle=-90}}

\centering{\ \hspace*{-4mm} \epsfig{figure=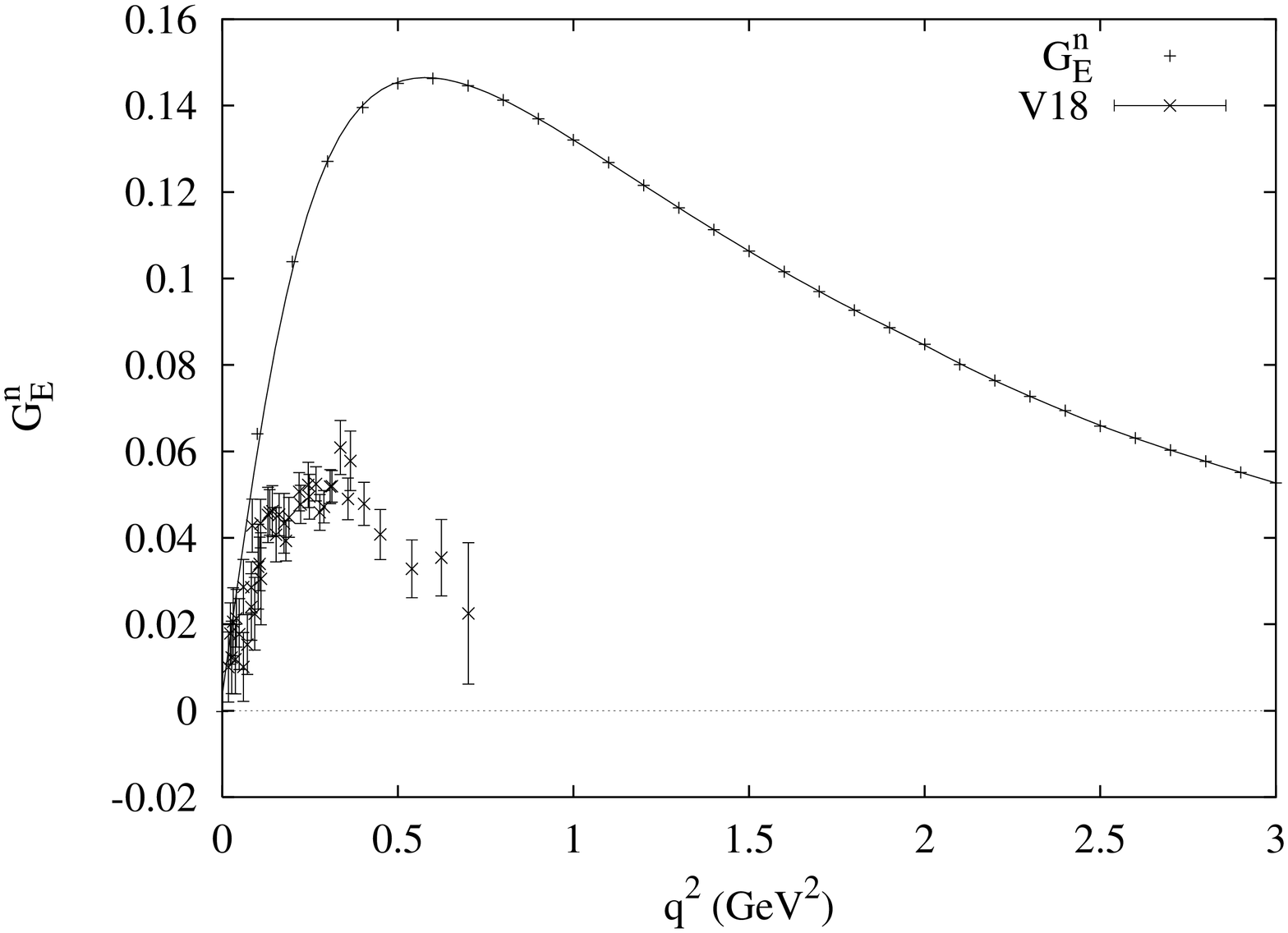,height=8.2cm,angle=-90}}

\caption{Upper panel: Calculated proton electric form factor: $+$,
compared with the empirical dipole fit: $F_{\rm emp}(q^2)= 1/(1+q^2/m^2_{\rm
emp})^2$, $m_{\rm emp}=0.84\,$GeV.  Lower panel: Calculated neutron electric
form factor: $+$, compared with the experimental
data~\protect\cite{saclay} as extracted using the Argonne V18
potential~\protect\cite{bobpot}.  In both calculations the Monte-Carlo errors
are smaller than the symbols.
\label{pnFF}}
\end{figure}

\begin{figure}[t]
\centering{\ \epsfig{figure=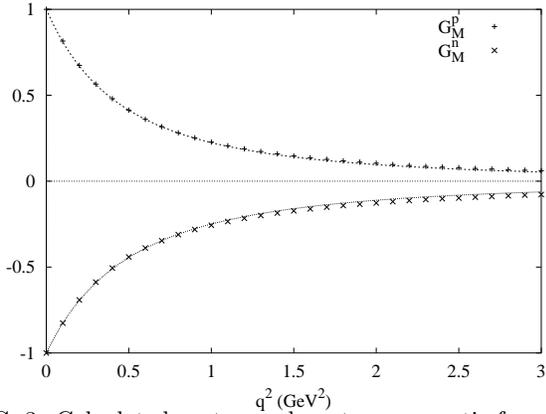,height=8.0cm,angle=-90}}
\caption{Calculated proton and neutron magnetic form factors, normalised by
$|\mu_{p,n}|$ in (\protect\ref{results}).  The curves are dipole fits with
masses (in GeV): $m_p=0.95$, $m_n=1.0$, $13$\% and $19$\% larger than $m_{\rm
emp}$ in Fig.~\protect\ref{pnFF}.  ($\mu^{\rm emp}_{p,n}\,F_{\rm emp}(q^2)$
describes the data very well.)
\label{magFF}}
\end{figure}

The calculated form factors are depicted in Figs.~\ref{pnFF} and~\ref{magFF}
and it is obvious in Fig.~\ref{pnFF} that we used $G_E^p(q^2)$ to constrain
our fit.  The $0^+$ $(ud)$ diquark correlation in $\Psi$ ensures that
$G_{E\,{\rm fit}}^n(q^2) \not\equiv 0$, and the presence of diquark
correlations can also explain the $N$-$\Delta$ mass difference.  Our result
for $G_E^n(q^2)$ is well described by~\cite{saclay}
\begin{equation}
G_{E\,{\rm fit}}^n(q^2)=
- \mu_n^{\rm emp}\,F_{\rm emp}(q^2)\,\frac{a^2\,\tau}{1+ b^2\,\tau}\,,
\end{equation}
with $\tau = q^2/(2 M)^2$, $F_{\rm emp}(q^2)$ given in Fig.~\ref{pnFF}, and
$a= 1.33$, $b=1.00$, and the discrepancy between our calculation and
experiment can be discussed in terms of these parameters.  $a$ characterises
the charge radius and it is $\lsim 30$\% too large, as can be anticipated
from (\ref{results}).  $b$ describes the magnitude at intermediate momenta
and it is only $\sim 23$-$35$\% of the empirical value.  That is a systematic
defect shared by other studies~\cite{mishakeinerraC} that only retain the
scalar diquark correlation.  Unlike those studies, however, our calculated
magnetic form factors, Fig~\ref{magFF}, agree well with the data and, as we
have seen, that is because we include the diquark breakup diagrams.   
\begin{table}[b]
\caption{A variation of the model parameters: $\omega_\psi$, $\omega_\Gamma$
and $m_\Delta$ (in GeV) illustrates the sensitivity and stability of our
results.  The column labelled ``$r_n$'' lists:
${\rm sign}(r_n^2)\,|r_n^2|^{1/2}$.  (Radii in fm, magnetic moments in units
of $\mu_N$.  The statistical errors are $\leq 1$\%.)
\label{sens}}
\[
\begin{array}{ccc|ccccc}\hline
\omega_\psi & \omega_\Gamma & m_\Delta & r_p & r_n & \mu_p & \mu_n &
\mu_n/\mu_p\\\hline 
0.20 & 1.0 & 0.63 & 0.79 & -0.43 & 2.88&-1.58 & -0.55\\\hline
0.16& 1.0& 0.63& 0.84 & -0.46 & 2.83 & -1.55 & -0.55\\
0.24& 1.0& 0.62& 0.75 & -0.41 & 2.89 & -1.59 & -0.55\\\hline
0.20& 0.8& 0.62& 0.80 & -0.40 & 2.93 & -1.64 & -0.56 \\
0.20& 1.2 & 0.63& 0.78 & -0.45 & 2.84 & -1.54 & -0.54 \\\hline
\end{array}
\]
\end{table}
\hspace*{-\parindent}It must be borne in mind that in our calculation $a$ and
$b$ are {\it not} independent.  Modifying the parameters in
(\protect\ref{params}) so as to reduce $a$ automatically and substantially
increases $b$.  However, not withstanding our observation that its importance
has previously been overestimated, without an axial-vector diquark
correlation it is not possible to accurately describe all observables
simultaneously.

%
We have employed a three-parameter model of the nucleon's Fadde'ev amplitude,
$\Psi$, to calculate an impulse approximation to the electromagnetic form
factors.  $\Psi$ represents the nucleon as a bound state of a confined quark
and confined, nonpointlike scalar diquark, and the exchange of roles between
the dormant and diquark-participant quarks is an integral feature.  Five
processes contribute: direct quark-photon scattering with a spectator
diquark; photon-diquark scattering with a spectator quark; and three distinct
diquark breakup diagrams.  We obtain a good description of all form factors
except $G_E^n$, which is too large in magnitude.  That defect is shared by
all models that do not include more than a scalar diquark correlation.  The
nonpointlike nature of the diquark correlation is important, especially via
the breakup contributions which provide large contributions to the magnetic
moments and ensure $\mu_n/\mu_p < -0.5$.

Including a nonpointlike axial-vector diquark is an obvious improvement of
the model.  That must be done in analogy with the scalar diquark because an
accurate interpretation of the model parameters is impossible if the breakup
diagrams are discarded.  Another avenue for improvement is a direct solution
of the Fadde'ev equation, retaining the axial-vector correlation and the
breakup contributions to the form factor.  That would provide a model for
correlating meson and baryon observables in terms of very few parameters.

Models of the nucleon such as ours have hitherto been applied only at small-
and intermediate-$q^2$.  Based on the observation~\cite{mr98} that a
description of the large-$q^2$ behaviour of $F_\pi(q^2)$ is only possible if
the subleading pseudovector components of the pion's Bethe-Salpeter amplitude
are retained, we anticipate that a successful description of the nucleon form
factors on that domain will require a parametrisation of the Fadde'ev
amplitude that includes the analogous subleading Dirac components.

\begin{table}[t]
\caption{Relative contribution to the charge radii and magnetic moments of
each of the five diagrams in our impulse approximation:
Fig.~(\protect\ref{ff}),
(\protect\ref{L1})--(\protect\ref{L5}). \label{ff1to5}}
\[
\begin{array}{r|rrrrr}\hline
{\rm diagram} & 1 & 2 & 3 & 4 & 5 \\\hline
(r_p^2)^i/r_p^2
 &\; 0.68 & 0.11 & -0.02 & 0.12 & 0.12 \\
(r_n^2)^i/r_n^2
 &\;   1.14 & -0.37 & -0.15 & 0.19 & 0.19 \\
\mu_p^i/\mu_p 
 &\;  0.60 & 0.01 & 0.04 & 0.17 & 0.17 \\
\mu_n^i/\mu_n 
 &\;  0.55 & -0.02 & 0.15 & 0.16 & 0.16 \\\hline
\end{array}
\]
\end{table}
%

This work was supported by the US Department of Energy, Nuclear Physics
Division, under contract number W-31-109-ENG-38.  S.M.S. is a F.-Lynen Fellow
of the A.v. Humboldt foundation.


\end{document}